# Astrophysicists and physicists as creators of ArXiv-based commenting resources for their research communities. An initial survey

Monica Marra

*INAF – Osservatorio Astronomico di Bologna, Via Piero Gobetti 93/3, I-40129 Bologna, Italy*
*E-mail: monica.marra@oabo.inaf.it*

**Abstract.** This paper conveys the outcomes of what results to be the first, though initial, overview of commenting platforms and related 2.0 resources born within and for the astrophysical community (2004–2016). Experiences were added, mainly in the physics domain, for a total of twenty-two major items, including four epijournals – and four supplementary resources, thus casting some light onto an unexpected richness and consonance of endeavours.

These experiences rest almost entirely on the contents of the database ArXiv, which adds to its merits that of potentially setting the grounds for web 2.0 resources, and research behaviours, to be explored.

Most of the experiences retrieved are UK- and US-based, but the resulting picture is international, as various European countries, China and Australia have been actively involved. Final remarks about creation patterns and outcome of these resources are outlined. The results integrate the previous studies according to which the web 2.0 is presently of limited use for communication in astrophysics and vouch for a role of researchers in the shaping of their own professional communication tools that is greater than expected. Collaterally, some aspects of ArXiv's recent pathway towards partial inclusion of web 2.0 features are touched upon. Further investigation is hoped for.

## 1. Introduction

Significant literature has proved that scholarly communities shape their online communication and information practices in different modes according to the different disciplinary domains ([15], with a review of the literature; [16]). These variably adjusted practices include uptake and use patterns of specific online tools or families of tools, as it has been illustrated, e.g., for social media within research environments [45,51,53].

The community of astrophysicists and that of physicists have received considerable attention with regard to these topics, probably for having pioneeringly taken the path of remodelling a significant part of their internal communication by means of the Internet, and conceivably – with specific regard to astrophysics – also for being a relatively small and tendentially self-contained scholarly community. Studies have ranked physics as the third discipline by use of social media in general [9,51], although appropriate warnings have been issued with regard to the specific behaviour of sub-disciplines [46]. As for astrophysics, the preferred modes of scholarly interaction have been convincingly found to consist of email exchanges and colloquia within working groups [25]. Also, it has been maintained that "astrophysicists have limited engagement with Web 2.0 technologies", while the role of "email networks" for communication has been stressed, in an overall setting where "face-to-face interaction remains an essential part of the collaborative process" [25]. This has later been confirmed by [13,33] (the former in a specific context, the latter for high energy physicists).

In recent times, though, a non-negligible diffusion of some 2.0 communication tools has in fact been detected. Light has been cast on the use of Twitter [26,27,29] and on that of professional social networks ([33]; at present, anyway, these practices don't appear to have decisively undermined the most established communication trends in the discipline.

The present research is aimed at understanding astrophysicists' and physicists' disposition towards paper commenting and rating in online contexts, through dedicated platforms. Interactivity is among the major marks of the web 2.0 era [40], and it interlinks with the progressive erosion of scholarly consensus around the classic form of peer-review ([23,54]; a review is in [3]). As such, scholarly commenting has been object of dedicated studies since the first decade of the present century. Neylon and Wu [37] have provided valuable considerations and insights into this practice at large: "commenting in the scientific community simply hasn't worked, at least not generally", because scientists "are used to criticizing articles in the privacy of offices and local journal clubs, not in a public, archived forum", which may damage careers; anonimity has pros and cons, namely it "can support more direct and honest discussion but [. . .] often degrades discussions [. . .]. Another issue is that the majority of people making hiring and granting decisions do not consider commenting a valuable contribution" [37].

In 2008, Michael Nielsen had supported the same perspective on his blog [39], and subsequently identified a further obstacle in researchers' tendency not to build the tools for online commenting on themselves, which would be a driver of success for this kind of practice. In his words, "to create an open scientific culture that embraces new online tools, two [. . .] tasks must be achieved: first, build superb online tools; and second, cause the cultural changes necessary for those tools to be accepted. [. . .] [The former] requires a rare combination of strong design and technical skills, and a deep understanding of how science works. The difficulty is [that] the people who best understand how science works are scientists themselves, yet building such tools is not something scientists are typically encouraged [. . .] to do" [38]. Procter *et al.* are in line, confirming that "so far [. . .] providing ratings or comments on articles has not proved popular" in various scientific contexts [45].

The research that follows proposes a contextual revision of some of these otherwise sharable considerations, as it tracks remarkable exceptions in scientists' behaviour towards the creation and – to some extent – uptake of commenting tools based on the world-famous preprint database ArXiv. The community of astrophysicists and – to the extent to which it has been surveyed – that of physicists, with the addition of some mathematicians, shows over time a persisting effort to create 2.0 tools of its own, destinated to colleagues, with the purpose of openly commenting papers posted on ArXiv.

## 2. Methodology

The present research can be estimated to have required about fifteen months of activity (FTE). The first documentation (both literature and web resources) was retrieved in late 2014, the last one in March 2017, with updates in Autumn.

Internet search engines have proved to be of limited usefulness in order to let these resources come to light. Queries have been executed with the phrases "arxiv comment*", "arxiv discuss*" and "arxiv peer review*". The first three pages of results (30 items) for these queries yielded only 24,93% of the twenty-two main resources here described (including the resources here simply mentioned would have lowered the percentage furtherly). Instead, important sources have been previous online compilations such as the wide shared database *400 + Tools and innovations in scholarly communication* (http://bit.ly/innoscholcomm-list, last visited March 20, 2017), first published in March 2015 by Bianca Kramer and Jeroen Bosman of Utrecht University Library and then constantly updated [31]. As at March 21, 2017, it listed as many as 668 resources. This unique collection has been thoroughly consulted in Spring 2016, with subsequent inspections later in 2016 and in 2017; at March 2017, it contained 31,81% of the resources in the main group, only one of which – *ViXra* – could be retrieved also through the search engine above. The utility of this resource has been concrete and unquestionable; anyway, due both to its continuous update and to the prolonged and multiple-source documentation activity needed to get to the present survey, it would be difficult to reconstruct exactly, and retrospectively, the percentage of *400 + tools* which represented an actual source for the present findings.

Particularly fruitful has proved to be tracking social media mentions, with special referral to blog comments ("snowballing"). These strategies have demonstrated a special effectiveness, as they made it possible to retrieve as much as 54,54% of the resources in the main list, plus one of those simply mentioned. A conversation with an astrophysicist was the original source for a further platform, *Cosmocoffee*.

Precious details about some of the projects surveyed came from email exchanges with some of the researchers involved, as will be detailed below (see also the acknowledgements).

About 60% of the bibliographic references was found in 2016, with a further ~30% being filed between 2013 and 2015. In fact, though, the literature was more useful for giving a profile to some of the themes involved than for providing concrete examples that be useful to the building of this survey. Actually, the literature was the original source for only two of the resources retrieved (*Naboj* and *The RIOJA Project*) – although some more of these resources have received attention by researchers, journalists or bloggers (some references are in the reference list).

The relatively long period of time which has been necessary to build up the present research has provided the possibility to follow-up some of the resources retrieved, checking for their persistence and – in a few cases – for the response from users over time.

The criteria for selecting the resources in this survey were: (a) having been created by researchers, (b) for their same scholarly communities, and (c) relying on ArXiv contents entirely or almost entirely.

## 3. The importance of ArXiv beyond preprints provision

The creation of ArXiv, the first and foremost preprint server in 1991, has been recognized as "the most significant change in scientific communication since the establishment of the journal in the 17th century" [17]. The importance of this novel way of circulating scientific papers exceeds that of enhancing papers' availability in a peculiarly early stage of their customary disclosure, as Arxiv has pioneeringly explored all the main changes in XX and early XXI century's scholarly communication practices, among which the progressive maturation and diffusion of the open access movement. The latter found ArXiv giving researchers the opportunity to upload accepted or published versions of papers, thus putting those principles into practice for the communities involved, while the massive hosting of preprint papers let this database be perceived as an implemented source for open contents, in spite of the conceptually specific nature of this task.

In fact, ArXiv's fruitfulness went beyond. As early as in 1994 – two years before it's often stated to have happened – Paul Ginsparg himself envisioned the possibility for ArXiv to act as a starting platform for add-on tools fostering not only dissemination but also validation practices, the latter through the birth of a network-based scholarly interactivity centered on the ArXiv eprints [21,22]. The classic article by Rodriguez et al. shows how cleverly these suggestions could be seized and developed just after the landmarking debut of the web 2.0 around 2005 [50]. Later on, further scholars highlighted the existence of potentialities for ArXiv with respect to 2.0 scholarly validation [1,18]. Meanwhile, as noted before, the traditional features and role of peer-review within the science production chain was increasingly questioned, while the milestone phenomenon of web 2.0 slowly began transforming academic practices – as acknowledged even in cautious scholarly perspectives [36]. Useful studies have aimed at tracking this process and at casting light on a variety of 2.0 tools for the scholarly communities, as well as on patterns of their use [2,12,14,31,51,53].

Nevertheless, the exact role of ArXiv within this global, substantial paradigm change doesn't result to have been fully investigated yet. Also, to the best of our knowledge there aren't any comprehensive studies about how the web 2.0 attitude has progressively affected the astrophysical field, in addition to the small amount of literature mentioned above – and below, the latter about some particular aspects.

Polydoratou and Moyle have interestingly surveyed astrophysicists' attitudes towards ArXiv overlay journals [42], in the context of a specific project accounted for in 4.a.3. *infra*.

As noted earlier, the use of Twitter among astrophysicists has received considerable attention in recent years; the conclusions seem anyway to downplay its role for internal scholarly communication, although from the present point of view it has been interestingly noted that "most tweets refer to the ArXiv instead of the publishers' versions" [27].

Ritson has examined some major socio-scientific aspects of the "trackback system" connecting ArXiv papers and scientists' blogs since 2005, with an account of the previous science-and-technology-studies (STS) literature on the subject [49]. From the present perspective, three points result to be fruitful: (a) blogs, although peculiar in type, may well be considered means for providing papers with scientific feedback, included peer-review; (b) in 2006, one year after the debut of the trackback system, blogs ArXiv had approved for trackback were 51 and trackbacks were 5132. If considered that (c) the high-energy physics community has long been discussing in order to find consensus on how to practically identify members enabled to have their blogs trackbacked to the ArXiv, these numbers cast light on a phenomenon that may well be considered potentially wider and significant.

Within the scientific communities, the topic of providing ArXiv with validation peer-review type capacities – or not – has long been debated, as researchers' blogs and forums can witness. An almost randomatic sampling – including the threads https://www.physicsforums.com/threads/a-peer-review-system-for-the-arxiv.568276/ (2012; last visited November 2, 2017) and http://academia.stackexchange.com/questions/32367/why-doesnt-arxiv-have-a-comment-section (2014, last visited November 2, 2017) – may provide an interesting insight into the views of shrewd and lively scientific communities.

## 3.1. *ArXiv and its present situation with respect to the web 2.0 setting*

It may appear somehow paradoxical that ArXiv, whose creator had so impressively timely envisioned his database's potential in the future web 2.0 ecosystem, hasn't been equipped with corresponding tools so far – notwithstanding ArXiv's persisting role as a pillar resource for astrophysicists. Paul Ginsparg's explanation for this slow pace has been the database's organizational framework due to budget and personnel constraints [24]. Things might now be changing to some extent, as in April 2016 ArXiv conducted an online survey among its users in order to "improve arXiv and think of future directions for the service" [48]. Meaningfully, opinions were polled about the possibility of adding a rating and an annotating system, thus allowing readers to comment on eprints on the database itself. In both cases, the respondents were almost equally divided between strongly in support – which was closely related to lower age – and strongly unfavourable (35% vs. 34% respectively, as for annotation; exactly 36% respectively, for the rating system: [48]), which appears to have led the team to not prioritize these issues. Anyway, the availability of a "next-generation arXiv" in three years' time [47] seems to be setting the technical grounds for some 2.0 services to be more easily provided in the future.

The present situation of ArXiv, jointly with its persisting overall popularity confirmed by the 2016 survey – 52.92% "very satisfied"; 42.43% "satisfied" – could bring to the supposition that a limited web 2.0 evolution of the database goes well with the astrophysical community's still prevailing inclination to tendentially preserve its current scholarly practices. In fact, there is significant evidence of commenting practices to ArXiv papers much beyond the traditional channels, with proper involvement of the web 2.0 setting.

## 4. Commenting on ArXiv

A largely practised mode of online interaction is represented by researchers' blogs and forums, which may comment on ArXiv papers. This specific channel is being barely mentioned here, as the complexity of the scenario and the relations with ArXiv through the so-called "trackback system" [49] would require an extended analysis. A single, early experience will anyway be accounted for and it's the one of *Cosmocoffee* (http://cosmocoffee.info/, last visited October 31, 2017), an astrophysical forum born in September 2004 as "intended for authorised arxiv authors and students" (http://cosmocoffee.info/faq.php). 2873 international users have registered as at October 31, 2017 (http://cosmocoffee.info/index.php). Although Cosmocoffee results to be a multi-purpose information resource, founders "hope that it can also become a useful reference resource, complementing the arxiv itself. [ . . .] Daily we discuss work and new papers with colleagues, either at our local coffee break or via email with colleagues all over the world. This discussion can be an extremely effective way to understand things better. As such, it seemed to make sense that those discussions be shared with others and be public. [ . . .] Therefore we set up cosmocoffee.info [. . .]" (http://cosmocoffee.info/faq.php#0). Posts can be read freely, but posting is only for the registered users. The sub-forum "ArXiv papers" appears to have started with a post by UK cosmologist Antony Lewis on September 24, 2004; last post was issued on August 15, 2014 (as at October 31, 2017) after a total of 1031 posts on 260 topics (http://cosmocoffee.info/index.php; other sections are still current). *Cosmocoffee*'s administrators result to be Sarah Bridle (University of Manchester), Olivier Dore (JPL-CalTech), Antony Lewis (University of Sussex) and Mike Nolta (Canadian Institute for Theoretical Astrophysics) (http://cosmocoffee.info/faq.php#0; affiliations as at present). For as much as it results, Cosmocoffee has never been object of dedicated studies.

The present survey will focus on different-type 2.0 resources which offer commenting features in the physical and astrophysical domains. For presentation purposes, it seems possible to roughly divide the resources retrieved into three main categories:

(a) Resources or projects aimed at buiding new, open access and more interactive forms of the traditional scholarly journals. The model is that of the "overlay journal" or "epijournal" [6,44,52];
(b) "Actual" commenting platforms;
(c) Different tools which can very roughly be defined as variant forms of ArXiv – with whom they have no kind of affiliation or other apparent link. These will be conveyed firstly, due to their peculiar characteristics. The tools in this section often have more limited web 2.0 capacities and are considerably different both from each other and partly from ArXiv, too. They witness a widespread effort to build upon the model, as well as ArXiv's totemic standing within the physics and astrophysics environments – e.g. in the names and graphic look of 4.c.2. and 4.c.3.

The resources will be described synthetically; for some more details, as at March 2017, see [34].

*4.c. "Variant" forms of ArXiv*

The definition, as said before, is intentionally broad and pragmatic, in order to group together online entities with commenting feature appearing to be, overall, secondary. Their focus, in fact, seems to be about modifying some of ArXiv's main features: either improving search functions, or renewing visualization features, or being suitable for a different audience, or changing authors' admission policy.

**4.c.1.** *ArXivsorter* (http://www.arxivsorter.org, last visited October 30, 2017) was born thanks to the young French astrophysicist Brice Ménard, and to Jean-Philippe Magué; it was registered on Sourceforge on July 30, 2007 (https://sourceforge.net/projects/arxivsorter/). On the homepage: "Arxivsorter uses the network of co-authorship to estimate a proximity between people. It then ranks a list of publications using a friends-of-friends algorithm", seeming thus to be aimed at customizing ArXiv-astrophysics papers' sorting for the registered users. "The Arxivsorter algorithm [ . . .] ranks a list of papers, without any loss of information" (http://www.cita.utoronto.ca/~menard/Arxivsorter_Documentation/).

**4.c.2.** *ViXra* (2009; http://vixra.org/, last visited October 31, 2017), created by the independent UK-based physicist Philip Gibbs, "has been founded by scientists who find they are unable to submit their articles to arXiv.org because of Cornell University's policy of endorsements and moderation.[. .]. ViXra is an open repository for new scientific articles. It does not endorse e-prints accepted on its website, neither does it review them against criteria such as correctness or author's credentials" (http://vixra. org/). Commenting presently happens outside ViXra, apparently mainly through the resource's account on Freeforums (vixra.freeforums.org). ViXra hosted 20700 outputs at October 31, 2017, 1378 of which in astrophysics (http://vixra.org/), and is equipped with active social media accounts. About ViXra, see [7] and [19]; the tool is listed in [31].

**4.c.3.** *SnarXiv* (http://snarxiv.org/, last visited October 31, 2017) was born in 2010 by initiative of David Simmons-Duffin, by that time a PhD student in high-energy physics at Harvard University, and in a somehow bohemian spirit. "The snarXiv is a random high-energy theory paper generator" (http://davidsd.org/2010/03/the-snarxiv/) – basically a parody. It includes an interactive game: "arXiv vs. snarXiv" (http://snarxiv.org/vs-arxiv/), where players have to spot genuine ArXiv titles from SnarXiv ones, and get rated for their performance.

**4.c.4.** *Astrobites* (https://astrobites.org/, last visited October 31, 2017) is a successful project created in 2010 by graduate students in astronomy. Its "goal is to present one interesting paper per day in a brief format that is accessible to undergraduate students in the physical sciences" (https://astrobites.org/about/) – although it's also a web portal for different-type information. Typically, the papers suggested are from ArXiv's astrophysics section "astro-ph". The resource is presently written by thirty people, mainly based in the US and in the UK (https://astrobites.org/meet-the-authors/). Past web hosting was at Harvard University, with the help of James Guillochon (https://astrobites.org/about/; see below, 4.b.3.); remarkably, "since 2016 Astrobites has been hosted and supported by the American Astronomical Society". Links to other commenting resources in astrophysics are provided (*VoxCharta*, *ArXiver*, see below); Astrobites has popular profiles on Twitter and on Facebook.

**4.c.5.** 2013 saw the debut of the impressing *PaperScape* (http://paperscape.org/, last visited March 17, 2017), "an interactive map that visualises the arXiv" in the form of a multi-coloured galaxy, with each star representing an ArXiv paper. The "map [. . .] can be explored by panning and zooming. The papers are sized according to their number of citations and positioned according to their references/citations. Different categories of the arXiv are assigned different colours, and newer papers are more brightly coloured. The original project complements this map by letting you draw graphs of the papers that interest you, with the papers as nodes and citations as links. It's possible to register a personal profile, with which you can tag relevant papers as well as save and share the graphs you make" [30]. Developers are young physicists Damien P. George, currently working at the Department of Applied Mathematics of the University of Cambridge, and Robert J. Knegjens. It was accounted for by several blogposts (http://blog.paperscape.org/?page_id=101).

**4.c.6.** October 2013 saw the debut of *arXiver* (http://arxiver.net/, last visited November 4, 2017), whose "original credit for the idea" is acknowledged to the young British astrophysicist and web 2.0 activist Robert Simpson; (co-)maintainers are the Australian postdoctoral student Vanessa Moss (CAAS-TRO: http://www.caastro.org/people/moss-dr-vanessa, visited November 5, 2017), and Aidan Hotan (http://www.arxiver.net/about/). While ArXiv is said to be highly appreciated, it is also maintained that it presently appears "not very nice to look at (too much text!)" and "it would be nice to be able to glance at a visually-appealing summary of different papers to then go forth and read properly" (https://arxiver.wordpress.com/about/); this seems to basically consist in providing selected pictures from the article by the side of the ArXiv abstract. In fact, registered users can also assign "likes" to papers' posts. An interesting feature was the initial absence of author names in new papers' posts, in order to correct for any possible author bias (https://arxiver.wordpress.com/faq/). Since its debut, ArXiver was equipped with a Twitter feed, which had 1041 followers as at November 5, 2017.

**4.c.7.** *Cloudy Science* (https://cloudyscience.wordpress.com/, last visited November 5, 2017) was born presumably either in 2014 or shortly before, but "revived" in January 2015 "after a long period of stagnation" (https://cloudyscience.wordpress.com/updates/). It is defined as a "partner site" by ArXiver (http://arxiver.net/). "The goal of Cloudy Science is to present automatically generated wordclouds that give a researcher insight into the content of a paper, offering another way to quickly judge whether a paper might be [. . . ] relevant to them. It currently only focuses on arXiv's astro-ph" (https://cloudyscience.wordpress.com/about/). Registered readers can assign "likes" to single papers, but this feature appears to have been very scarcely used. At the moment of writing, Cloudy Science is "brought to you" by Vanessa Moss.

*ArXivist* (http://arxivist.com, last visited November 5, 2017) and *ArXiv Sanity Preserver* (http://www.arxiv-sanity.com, last visited November 5, 2017) were both born in 2016, the latter's interface appearing more sophisticated and appealing than the former as at writing. They also share the feature of using readers' preferences – as provided in a web 2.0 environment – for customizing ArXiv daily updates for users accordingly. Both developers (Anton Lukyanenko and Andrej Karpathy respectively) are US-based and are active in the mathematic field (the former) and in computer science (the latter), which suggests not to get into further details in the present context.

We're not going into other meaningful alike projects in the scientific domain either, partly as they appear to be multidisciplinary (such as *Academic Karma*), and because they don't show a tight link with ArXiv (e.g. *Preprints*, https://www.preprints.org/). This doesn't mean that some of them may not have reached interested astrophysicists and may have been explored to some extent.

*4.a. ArXiv-based overlay journals and projects*

Mathematicians, computer scientists and physicists have shown an active attitude at implementing ArXiv-based overlay journals ([6,44]; early examples in [28]). Meaningful samples of computer sci-entists' views on the subject, supplemented by a specific project, can be read at the blogpost "Scientific journals in the e-publishing age", written by computer scientist Philip Thrift on his blog "Occupy publishing" on February 1, 2012 and widely commented (http://occupypublishing.blogspot.it/2012/02/scientific-journals-in-e-publishing-age.html, last visited November 6, 2017). Also to 2012 but to the mathematic field seems to have belonged the project of *arXiv Review* (no more available at http://arxiv-review.org/ in March 2017), apparently intended as an ArXiv overlay journal with commenting and rating features. Related documentation can still be found e.g. at http://occupypublishing.blogspot.it/2012/02/guidelines-for-arxiv-review.html, last visited November 8, 2017. In the same domain, new projects have been implemented recently, such as Tim Gowers' *Discrete analysis* (http://discreteanalysisjournal.com/, 2015, last visited November 6, 2017; announcements on Gower's blog, e.g. https://gowers.wordpress.com/2016/03/01/discrete-analysis-launched/, last visited November 6, 2017).

New achievements have been accomplished in physics, too and will be accounted for in more detail.

**4.a.1.** Dutch platform *SciPost* (https://www.scipost.org/, last visited November 5, 2017), founded in 2016 by Jean-Sébastien Caux, professor of theoretical physics at the University of Amsterdam, presently provides three ArXiv-overlay publications: "SciPost Physics", "SciPost Physics Lecture Notes" and "SciPost Physics Proceedings" (https://www.scipost.org/journals/), whose contents are published under the CC-BY 4.0 license and equipped with DOIs (https://www.scipost.org/FAQ). Commenting is possible for registered contributors. "SciPost Physics", indexed in GoogleScholar, publishes research articles in experimental, theoretical and computational physics, including cosmology and astroparticle physics (https://www.scipost.org/10.21468/SciPostPhys/about); as at writing, seventy-one accepted articles have been published. Outstandingly, Scipost is endorsed by the Netherlands Organization for Scientific Research (NWO) (https://www.scipost.org/), which financed the startup phase (https://www.scipost.org/FAQ#scipost_funded). *SciPost* relies upon an international editorial college of about fifty members (as at November 6, 2017), with an advisory board of ten academics from the Netherlands as well as from other European countries. This resource is cited by [31].

**4.a.2.** *Quantum* (http://quantum-journal.org/, 2016, last visited November 6, 2017) is a further successful, open-access, not-for-profit, peer-reviewed ArXiv-overlay journal, sharing with *SciPost* some more features, and is active in quantum science and related fields. "All papers submitted to Quantum must be listed on (or cross-listed with) the arXiv section quant-ph. In case of acceptance, the final version must be uploaded to the arXiv before publication" (http://quantum-journal.org/about/faqpage/). Thirty-two articles have been published as at writing.

In an interview to the blog "Scholastica", co-founder Christian Gogolin (ICFO Institute of Photonic Sciences, Barcelona) states that "we were strongly inspired by other arXiv overlay journals; perhaps *Quantum's* distinguishing feature is the strong emphasis on community involvement" (http://buff.ly/2k5yqUx, last visited November 6, 2017). The fourty-members editorial board is mainly European. Accepted papers are published under a CC-BY 4.0 licence and receive a DOI through Crossref; "Quantum is backed by a democratic non-profit society" (the viennese Verein zur Förderung des Open Access Publizierens in den Quantenwissenschaften: http://quantum-journal.org/impressum/, last visited November 6, 2017). A subreddit has been provided for feedback and discussions, https://www.reddit.com/r/quantumjournal/ (last visited November 6, 2017); own Twitter and Facebook profiles look popular, the former resulting to have 1705 followers, the latter 1272 as at November 7, 2017.

In the field of astrophysics, a single example of ArXiv-based overlay journal has seen the light up to the moment (4.a.4., infra), but previous, sometimes advanced efforts in this direction had been made before.

In a blog comment to the later experience of 4.a.4. (infra), Daniel Fischer witnessed that about 1997 some researchers attending a conference in Germany had already conceived the idea of creating a journal "ArXiv mated with open peer review" [. . .] the name that journal should be given: "Open Astronomy", but "the concept never saw the light [. . .]". It seems credible that the same consideration has arisen elsewhere too in the global astrophysical community; this is proved as at June 2005 among a group of young but very mindful British astrophysicists contributing to CosmoCoffee, which included Antony Lewis and Sarah Bridle (http://cosmocoffee.info/viewtopic.php?t=276, last visited November 7, 2017).

Some years later, two relevant projects reached far more advanced, though different, stages of fulfilment and appear to be or have been well-rooted within the astrophysical community.

**4.a.3.** The first one was the impressing *RIOJA Project* (Repository Interface to Overlaid Journal Archives, 2007), which has been recognized as the first overlay project in astrophysics [41].

This initiative was supported by prominent scholarly institutions both in the UK and in the USA: University of Cambridge, Imperial College London, University of Glasgow, UCL, Cornell University, and funded by JISC. It was preceeded by a careful examination of the side conditions inclusive of a wide survey created by Polydoratou and Moyle [42,49] as well as by a feasibilty study [43]. A final report was also provided in 2008 [35]. Although a demonstrator implementation was achieved, as witnessed by the final report, it results that no overlay journal has subsequently been built on that technology. The RIOJA Project has been accounted for by [6,8,44].

**4.a.4.** Five years later (2012), and still in a UK context, a new project was launched by professor Peter Coles, theoretical cosmologist at the University of Cardiff, and eventually led to *The Open Journal* (http://astro.theoj.org, last visited November 7, 2017). The launch of the initiative came with the blogpost *A Modest Proposal – The Open Journal of Astrophysics*, published on Coles' blog "In the dark" on July 17, 2012 – following previous discussions within and outside this blog [10]. In Coles' words: "[. .] My suggestion is that we set up a quick-and-easy trial system to circumvent the traditional publishing route. The basic is that authors who submit papers to the arXiv can have their papers refereed by the community, outside the usual system of traditional journals. I'm thinking of a website on which authors would simply have to post their arXiv ID and a request for peer review. Once accepted, the author would be allowed to mark the arXiv posting as "refereed" and an electronic version would be made available for free on the website" (ibid.); the accepted articles are published under a CC-BY license and the reviewer comments can be disclosed "at the joint discretion of the authors and reviewers" (http://astro.theoj.org/about). Almost 70 qualified comments were received from other scholars within the following fortnight (plus others successively). Interestingly, one of them came from one of the researchers previously involved both in the mentioned discussion on CosmoCoffee in June 2005 and later in the RIOJA project, who is now a member of OJ's editorial board (http://astro.theoj.org/about). Also, Robert Simpson (see 4.c.6. above) collaborated to the code development (https://telescoper.wordpress.com/2014/05/10/the-open-journal-for-astrophysics-project/, visited November 7, 2017). On 22 December 2015 it was announced that "the Open Journal is open for submissions" at http://astro.theoj.org/ [11]; shortly after, *Nature* dedicated an article to the OJ [20]. As of November 7, 2017, three papers appear as "accepted".

### 4.b. "Actual" commenting platforms

**4.b.1.** *Naboj* (http://www.naboj.com/, last visited November 8, 2017) was created in 2004/2005 (http://www.naboj.com/news.php) and now appears to be abandoned. Its name seems to be an anagram of the first name of its creator, Bojan Tunguz, who reports to have been "an international [physics] student and faculty at various US colleges and universities" (http://www.tunguz.com/About/, last visited November 8, 2017). The tool is described as "a dynamical website that lets [ . . .] [registered users] review online scientific articles [ . . .] that have been posted at Los Alamos ArXiv and PubMed Central". In fact, the papers commented come almost exclusively from ArXiv.  The resource seems to have been used by a restricted number of physicists: from 2005 to 2010, 23 comments were made, almost entirely on physics papers; more than 78% were made during the first two years of Naboj's existence. Rather interestingly, comments themselves could be voted as "useful" or "not    useful". The last  review available is dated February 18,  2010 (http://www.naboj.com/recent_reviews.php).  Naboj was accounted for by [12] and [8], as well as mentioned in [54].

**4.b.2.** *Scirate* (https://scirate.com/about, last visited November 8, 2017) was originally created by US physicist  and computer scientist   Dave Bacon in January 2007 (http://scienceblogs.com/pontiff/2010/06/07/what-to-do-with-scirate/, last visited November 8, 2017), then rewritten by Bill Rosgen in early 2012 (https://groups.google.com/forum/#!topic/scirate/wnjkKSZYZkI, dated 24.4.2012). Its code is on GitHub and data are published under a CC-BY-SA license (https://scirate.com/about).    The information on its features appears to be synthetic on the website – at least for non-registered users ("Follow arXiv.org categories and see the highest ranked new papers; scite [i.e.: vote] papers and subscribe to categories, sign up to customize your view of the site" (*ibidem*), but the interface is rather self-explanatory. Presently, eighteen ArXiv categories are available, including astrophysics, various branches of physics, mathematics and computer science.  Users need to have registered.  For a temptative assessment of its usage, the ArXiv papers which had been "scirated" at least twice during the year from 6 April 2015 to 6 April 2016 were 53; 2 of them had been commented. The resource appears to be more widely used by mathematicians and physicists, which is probably related with what seems to be the predominant research interest within the *Scirate* working group ("the Scirate Collaboration", https://scirate.com/about), i.e. quantum physics. *Scirate* is listed in [31].

**4.b.3.** *VoxCharta* (http://harvard.voxcharta.org, 2009, last visited November 8, 2017) is somehow peculiar among the tools in this group, as it provides rating and commenting on ArXiv papers for a practical aim: selecting papers for subsequent real-life scholarly discussions. Thus, this resource seems to bridge the gap between the two different ecosystems of virtual and real-life scholarly communication, possibly making its adoption easier also by research communities which appreciate more traditional means of internal communication. At present, the homepage states that researchers from 419 worldwide research institutions have registered for using VoxCharta with their colleagues – although not      all of these communities are active (usage data at http://ucsc.voxcharta.org/tools/institution-stats/).VoxCharta is self-defined as "a clone of arXiv used primarily for astronomy and astrophysics paper discussions. Users have the ability to vote for papers they would like to talk about at the next local discussion session. All papers that received votes [ . . .] appear in an "agenda" at the top of the main page, sorted by the number of votes each paper receives [ . . .]. Additionally, each paper has a "comments" link that allows you to post things that people who are reading astro-ph may find interesting, or might be useful to look at when talking [ . . .] at a discussion section. Viewing the web page can be done anonymously, but voting and commenting on papers requires an account. [ . . .] each department that uses Vox Charta has a person designated as a "liaison" who approves all new accounts for that department" ( http://harvard.voxcharta.org/about/about-this-website/). Starting from a local account, you can use VoxCharta globally by checking the preferences given to a paper worldwide. VoxCharta was designed and is maintained by astrophysicist James Guillochon (*ibid*.), currently at the ITC within Harvard-Smithsonian Center for Astrophysics (https://itc.cfa.harvard.edu/people/james-guillochon).  Thanks to his courtesy,  we know that the first  discussion took place on July 28,   2009;  after a couple of months,   due to other institutions' expressions of interest, the ability for the site to support multiple institutions simultaneously was implemented. The original number of ArXiv categories was gradually extended including, e.g., high energy physics. VoxCharta is listed in [31].

**4.b.4.** Another prominent experience is *PaperRater* (http://www.paperrater.org/, last visited November 8, 2017), created by young German astrophysicist Peter Melchior in 2010. Differently from other platforms, "PaperRater.org reads the daily submission to any category of arXiv and [also] searches for published papers on The SAO/NASA Astrophysics Data System (ADS) [ . . .]", the latter notoriously being the main database for published articles in astrophysics. The tool's fundamentals are stated as follows: "You can help [. . .] by rating, tagging or commenting papers. You can rate every paper only once, but you can change the rating later at any time. Your rating is anonymous. The distribution of ratings will be shown once a sufficient number of ratings is reached. You can add as many tags to each paper as you like [. . .] No other user can find out, which papers you rated or even what your rating was, nor what tags you chose. In contrast, comments are meant to be public. If you [ . . .] decide that you want to stay anonymous [. . .] you can choose to do so for any comment independently" (http://www.paperrater.org/help/getting-started.html, dated March 3, 2012). PaperRater's interface looks user-friendly and the tool's mission is clearly stated in the first post of the dedicated blog (October 8, 2010): "The peer review process has a long-standing tradition in improving manuscript quality [ . . .] However, it is not infallible [. . .] as students and researchers we all read papers daily [ . . .] and judge them [. . .] this process is able to improve a paper's quality beyond what a single referee could achieve. If the joint wisdom of the community could be bundled. This is what PaperRater.org is all about: to augment and eventually replace the intransparent process of peer review [ . . .] by a public one" (http://www.paperrater.org/blog/mission-statement.html).

In March 2016 the author's kindness made it possible to give some figures of users' response over time. Reads had increased significantly from 2010 (1467) to 2012 (2964), starting then to decrease (678 in 2013) until the last year available (363 in 2015). Ratings had reached a maximum during the first year (111) and decreased markedly after 2013 (when they were 20). Registered users were 558 – as at March 20, 2016.

**4.b.5.** The idea of *YouASTRO* (http://youastro.dyndns.org:43905/, last visited March 17, 2017) came during a post-conference international evening colloquium among astrophysicists – as kindly reported from project co-creator, Italian dr. Fabrizio Bocchino (Italian National Institute for Astrophysics), who wrote the YouASTRO code. The other involved researchers were Javier Lòpez-Santiago, Juan F. Albacete-Colombo and Niccolò Bucciantini. The tool was operative in 2011 and the project was presented to that year's ADASS conference [4]. As the platform is no more reachable at its url as at writing, the subsequent information comes from a visit made on March 17, 2017, unless specified differently.

"YouASTRO is a web application which allows us to leave comments and give rating to refereed astrophysical papers. For now, the papers which can be commented are only the papers appearing on the SAO/NASA Astrophysics Data System [ . . .] The YouASTRO Board of Editors think that the YouASTRO "leave a comment" feature can be of great benefit to the scientific community, if used widespreadly. It promotes the online scientific discussion focussed on papers [ . . .] in the framework of a general and continuous improvement of the quality of scientific publications, and the overall advance of science" (http://youastro.dyndns.org/faqs.html). "Registered users can vote a paper, one vote per paper [ . . .] rating goes from 1 (very poor) to 10 (excellent). Ratings are always anonymous [ . . .] YouASTRO only shows average ratings [ . . .] after more than 3 ratings have been received". The focus results to be on published articles rather than on preprints (e.g.: "comments to astroph papers will be automatically migrated to the refereed version ( . . .) when it appears"). As at June 2016, YouASTRO had 434 registered users (were 100 on 20.12.2011, http://youastro.dyndns.org/news.html, visited July 4, 2016); peaks of activity were achieved during the first years of operativity. Public comments result to be only 34,69% and, among them, anonymity is the standard (92,08%).

**4.b.6.** *ArXaliv* was created presumably at the beginning of 2012 by young mathematician Ralph Furmaniak, who was a PhD student at Stanford University by that time. When publicising his platform on a forum for colleagues on March 28, 2012, Furmaniak wrote "I have set up the reddit software to work with the arxiv database [ . . .] Each day it will update the list with the latest papers and you can upvote, downvote, comment, save links of interest, search, post new links, or create your own communities/arxalivs to post in or have others post links or writings of interest to them. [ . . .]" (http://publishing. mathforge.org/discussion/83/, last visited November 9, 2017). Exactly one year later, Furmaniak posted ArXaliv's codebase on GitHub in case "one day [ . . .] there are other people interested" (https://github. com/rfurman/arxaliv, last visited November 9, 2017). In fact, the tool looked "defunct" to another mathematician on a blogpost dated November 12, 2013 and is presently no more available at the original website http://arxaliv.org/.

**4.b.7.** *Selected Papers* (https://selectedpapers.net/) was developed in 2013 by US computational biologist Christopher Lee (https://johncarlosbaez.wordpress.com/2013/06/07/the-selected-papers-network-part-1/, dated June 7, 2013, last visited November 9, 2017; see also Lee's blogpost https://johncarlosbaez. wordpress.com/2013/07/12/the-selected-papers-network-part-3/, and [32]); the project results to have been supported by US mathematical physicist John Carlos Baez. This tool, which enabled commenting on ArXiv papers, had distinctive features among which using Google+ authentication and seems to have raised interest among researchers. In March 2016, anyway, Selected Papers resulted to be unaccessible, as it's still at present (a detailed documentation about it is still available at http://docs.selectedpapers. net, last visited November 9, 2017). Due to this situation, and to Lee's specific research area, a more extensive account of this resource won't be provided.

**4.b.8.** *r/Xiv* (https://www.reddit.com/r/Xiv/, last visited November 9, 2017) is "an interdisciplinary reddit for discussing papers submitted to arXiv, an open-access journal for e-Prints." It "aims to support arXiv by providing an open forum for papers and by calling attention to great papers" (ibid.). Registered users – 435 as at November 9, 2017 – can submit text posts or arXiv abstracts, and may receive comments from other registered users. Deductively, r/Xiv made its debut in 2014. In March 2017, posts – which can be upvoted – resulted to be fourty-seven, fourty-one of which were made in 2014, two in 2016, four in 2017; 53,19% of them received one or more comments. 80,85% had a tag and these are in many subfields of physics, included astrophysics, though the great majority were in quantum physics. There are two moderators, who appear to be active in quantum physics; only their nicknames are available and apparently they can't be contacted by non-members.

It can be noted that Reddit hosts further relevant subreddits, e.g. in cosmology and in astronomy, but the discussions don't appear to be based upon ArXiv papers.

**4.b.9.** *ArXiv Analytics* (http://arxitics.com/, last visited November 10, 2017) was developed in 2014 and is maintained by Chinese graduate student on high energy physics Zan Pan (Institute of Theoretical Physics, Chinese Academy of Sciences of Beijing). Collaboration and feedbacks was gained also from other nations (https://github.com/arxitics/arxiv-analytics/network/members, last visited November 10, 2017).

This resource is defined as "a set of projects [ . .]. As a first project, [it is] a web portal that offers more features and a better user interface for reading eprints provided by arXiv.org. You can search, subscribe, bookmark, review eprints, and interact with the community. The project is still under development." (http://arxitics.com/site/about).

ArXiv Analytics' functions appear to be manifold, the main ones being: "advanced search interface to find articles" (includes sorting by "reader counts" or by "rating score"); configure eprint subscriptions – e.g. by keywords, tags, authors; manage one's preferences/activities in a personal account (e.g. bookmarks, reading, rates, votes); post reviews and make comments; upload one's original content that has not been published online (under CC BY-SA 4.0 license; all data from http://arxitics.com/, visited November 10, 2017), thus gaining twenty "reputation points" for each document (http://arxitics.com/help/documents). The reputation system (http://arxitics.com/help/reputation) shows some apparent oddity such as losing reputation points when rating an article or voting a review ( $-1$ in each case, but $+5$ for publishing a new review); this is probably due to a value system that encourages sharing significantly ($+20$ for sharing a document) rather than judging on a small scale.

Thanks to Zan Pan's courtesy we get to know that there were 295 registered users as at March 2017, many of which are Chinese students; for them, *ArXiv Analytics* also provides a chat. The number of rated papers is presently "less than 100" (the feature is still experimental). Users, including guest ones, have been estimated to be "about 29.000 in total".

**4.b.10.** Another tool which appears to have been tailored upon ArXiv in a web 2.0 environment was *ArQuiv* (http://arquiv.org), which was presumably born in 2014. It was retrieved and visible on March 23, 2016 but no more available in March 2017, which is unchanged as at writing. Anyway, even more than it happens with other similar tools, the information supplied on the website was poor for those not registered, so that for example it was impossible to credit ArQuiv to its authors from the outside otherwise than "arQiv.org has no affiliation with arXiv.org or Cornell University" – and the homepage description was limited to: "arQiv.org: revolutionize scientific discussion by connecting readers and authors. To discuss any arXiv article, just change "**X**" to "**Q**" to visit arQiv". One of the ideas seemed to be to modify the typical url of an ArXiv paper in order to enable commenting.

**4.b.11.** *Benty Fields* was created in 2015 by young physicist Florian Beutler and cosmologist Morag Scrimgeour (http://www.benty-fields.com/, last visited November 11, 2017). On the homepage, the resource is described as "the academic network with daily papers and journal club organizer". In fact it's more than this, as it allows registered users to read ArXiv papers, possibly through recommendations built upon a machine learning algorithm trained by users' past preferences; save papers in various categories through the "library option"; "vote for papers, to put them on the next journal club agenda; leave comments or questions for papers; [. . .] start a journal club;" when using the search function, leave comments, vote for the paper, add it to the library or recommend it to a colleague. All the functions, as shown on the YouTube videos provided, offer manifold options. The accurate "job market" section can be supplemented with deadline reminders (http://www.benty-fields.com/). A notable characteristic is the tool's social networking feature. The interface is agreeable and the tool is sophisticated enough to provide a section about Terms and conditions as well as a privacy policy (http://www.benty-fields.com/tos#priv).

## 5. Conclusions

The availability of an established and comprehensive database of open access literature in physics and astrophysics such as ArXiv is likely to have fostered the birth of a significant number of web 2.0 experiences in these research fields and may have shaped them as electively literature-based. This seems to have happened rather early in some cases and anyway independently from ArXiv's adoption of a web 2.0 setting. In this respect, the vision of ArXiv as a founding ground for physicists' accreditation within their community results to be appropriate, not so much as the elegant proposal of a database having a legitimizing role for itself among physics researchers [13], but rather as an actual catalyst for web 2.0 scholarly exchanges within astrophysics and physics.

On the basis of the 2016 users survey, the ArXiv team appears now to be somehow mediating between researchers' "conservative" and still prevailing attitude, focussed on keeping the platform "to the core mission", and an emerging 2.0 trend which favours innovations such as rating and commenting on top of it. The ArXiv-Next Generation initiative, whose development has only just started [47], might perhaps mark the beginning of a change in this respect, for as much as it's possible to understand at present.

As for the tools here surveyed, and again for as much as it has been possible to observe, their outcomes appear to have been often affected by the physical limits of the local circles involved. For example, it has been found repeatedly that researchers committed to a project didn't know about the existence of parallel efforts among other colleagues, or that the news about a project's development didn't circulate well enough among interested people outside the circles – as witnessed by blog comments. An apparently rare piece of research about extending ArXiv's features to open peer-review and publishing [5] doesn't mention any of the ArXiv-based commenting resources for scholars which were already in place by that time according to our findings. All this testifies that, although obviously Internet-based, many of these experiences were in fact very local level-dependent, at least during the first years of their existence. All in all, actually, web 2.0 tools in astrophysics seem to have been strongly affected by local circumstances, both for the good – e.g. motivation – and for the bad. In the latter category a first-rate role should be acknowledged to the fact that restricted scholarly communities can seldom provide the critical mass for a new tool to take off, especially when validation features are involved. Shareably, the literature has remarked the crucial role of the critical mass as an essential driver for innovation in scholarly communication practices (e.g.: [37,55]). Anyway, for those resources for which the data available seem to be sufficient, you can deduce that peaks of activity are reached approximately in the two years after the debut. Platforms having prevailing functions other than paper commenting or rating show different workflows and life spans (e.g.: ViXra, CosmoCoffee).

For a significant part, the web 2.0 tools which have been accounted for above appear to have been created in a few astrophysical circles, mainly located in the UK and in the USA; specially lively environments have proven to be the University of Sussex and Harvard University. Following the academic pathway of some of the creators of these tools,    who sometimes were foreign students or researchers, might contribute to the history of web 2.0 commenting platforms in astrophysics.      This anyway goes beyond the aim of the present study and is probably more appropriate for retrospective future research. There are clues that   this aspect,  and the common local   perspective,  might  be changing in the latest years – approximately starting around 2012,   e.g. with a stronger presence of multi-national development teams. This might have to do with the diffusion of worldwide sharing platforms such as GitHub, although this is a simple hypothesis. 2012 also seems to be the peak of one of the time flows in which the experiences surveyed seem to have debuted – which is in line with Peter Melchior's remark as expressed in a comment to mathematician Philip Thrift's blogpost ("the internet seems to be bursting these days with ideas about how to improve/replace peer review and classical journal. This is a very exciting time. [. . .]", http://occupypublishing.blogspot.it/2012/02/scientific-journals-in-e-publishing-age.html, dated February 1, 2012; last visited November 12, 2017).

Other meaningful, though essential observations may concern how dedicated social media accounts have been created for some of the platforms and resources surveyed, and "coupled" to them. Sufficient, reliable information about this was retrieved for 63,63% of the 22 main resources. Among these, Twitter results to be the most used resource (45,15%; not all profiles are active), followed by Facebook (27,97%) and by blogs (15.25%). A provisional account of their use exceeds the limits of the present study. Also about intersections with social media, sharing buttons for posting papers to users' profiles have been provided by 11 out of the 14 resources for which information publicly available seemed to be sufficient. 100% of these resources provide sharing to users' profiles on Facebook; 81,81% to Twitter; 72,72% to Google+, plus a variety of others (Mendeley, Reddit, Pinterest et al.). It is notable that ArXiv has been providing the same service since April 2011 (https://arxiv.org/new#apr2011, for CiteUlike, Bibsonomy, Mendeley, Reddit, ScienceWise and other nonmainstream social resources). The sensation for these interlinks is that they aren't a factor of success for the platforms *per se*.

As a final remark, many of these resources clearly appear to have born in open-access-sensitive environments (particularly, but not only: 4.b.2., 4.b.3. 4.b.4., 4.b.9., all the epijournals), which flags socio-cultural motivations for their creation and is often an indicator for the presence of young personalities and groups.

## Acknowledgements

The author would like to thank for the information provided (in alphabetical order): Fabrizio Bocchino, Peter Coles, Christian Gogolin (in the name of the Executive Board of "Quantum"), James Guillochon, Antony Lewis, Peter Melchior, Martin Moyle, Zan Pan, Panayota Polydoratou, Oya Rieger.